\documentclass[12pt]{article}
\usepackage{amsmath}
\usepackage{graphicx}

\usepackage{epstopdf}

\usepackage{verbatim}

\begin{document}

\begin{center}

\Large{\bf Simultaneous determination of mass parameter\\and  radial marker in Schwarzschild geometry}

\vskip 1.0truecm

\renewcommand{\thefootnote}{\fnsymbol{footnote}}

\normalsize{Victor Varela\footnote{Previous institutional address: Institute of Mathematics, University of Aberdeen, King’s College, Aberdeen AB24 3UE, United Kingdom.}\\
victor.varela.abdn@gmail.com}

\vskip 0.5truecm

\normalsize{Lorenzo Leal\\
Centro de F{\'i}sica Te{\'o}rica y Computacional, Facultad de Ciencias,\\
Universidad Central de Venezuela, Caracas, Venezuela\\
lorenzolealb@gmail.com, lleal@fisica.ciens.ucv.ve}

\vskip 0.5truecm

\today

\end{center}\emph{}

\begin{abstract}
\noindent We show that mass parameter and radial coordinate values can be indirectly measured in thought experiments performed in Schwarzschild spacetime, without using the Newtonian limit of general relativity or approximations based on Euclidean geometry. Our approach involves different proper time quantifications as well as solutions to systems of algebraic equations, and aims to strengthen the conceptual independence of general relativity from Newtonian gravity.
\end{abstract}

\setcounter{footnote}{0}

\section{Introduction}
The static, spherically symmetric Schwarzschild metric is the most fundamental exact solution to Einstein's equations beyond flat spacetime. Its physical applications often require knowledge of the mass parameter $m$ and the radial marker $r$ of one or more events.

General relativity textbooks usually implement the Newtonian limit to identify $m$ with Newtonian gravitational mass $M$ multiplied by known constants. Specifically, it is noticed that the approximate, relativistic (geodesic) equation of motion of slow test particles in a weak, non-rotating gravitational field resembles Newton's equation of motion \cite{fn}. This suggests an asymptotic association of the Schwarzschild solution with the geometry of prerelativistic physics\footnote{See \cite{wld} for a comparison of basic geometric notions in Newtonian and Einsteinian physics.}. However, the Schwarzschild metric approaches the Minkowski line element in faraway spatial regions, so the light cone structure persists and the spacetime of Newtonian physics does not emerge as a limiting case. Certainly, such behaviour of the metric prevents the use of Newtonian coordinates (undergoing Galilean transformations) in the aforementioned regions.

The above considerations indicate that the link between $m$ and $M$ is not straightforward. Further elaboration is needed to grasp the underlying structure. Historically, that was achieved following the geometrisation of Newton's spacetime proposed by Cartan \cite{ec1,ec2} and Friedrichs \cite{fri} in the early stages of general relativity. After the contributions of several authors that spanned almost six decades (see \cite{daut} and references therein), Ehlers's presented his frame theory \cite{ej1,ej2} in which "light cones open up and become space-like hypersurfaces of constant absolute time" \cite{buma}. We emphasize that, regardless of the chosen implementation of the Newtonian limit, the appearance of $M$ in the Schwarzschild metric indicates some degree of dependence on Newtonian gravity and subjacent Euclidean geometry.

Practical determinations of the Schwarzschild coordinate $r$ also deserve careful attention. Radial marker, mass parameter and ruler distance are related through the spatial part of the metric. It has been pointed out that "ruler measurements are neither practicable nor relevant" in astronomical applications \cite{rnd}. So this connection may not be genuinely useful. Alternatively, the radar distance method \cite{rnd}, involving the round-trip travel time of a radio signal, could be implemented. If the radar distance between two events as well as $m$ are known, then the radial coordinates of the events are constrained. Thus, the radial marker of one event can be quantified only if the radial marker of the other event is known. Notably, the option of measuring radar distance with a clock placed at the centre of the material source of the gravitational field, and then equating the corresponding value of $r$ to zero is pointless\footnote{We implicitly assume that a non-rotating, static, spherically symmetric distribution of matter with an outside boundary constitutes the source of the Schwarzschild geometry defined in the outer, vacuum region. The inner, non-vacuum region features a different spacetime metric, which is generally unknown. Even if we could find the centre of an astronomically-sized material sphere and place a measuring instrument there; and deal with severe effects of the source material on the propagation of electromagnetic signals, knowledge of the coordinate radius of the outside boundary as well as the inner metric would be required to complete the gravitational model and determine radar distance. This observation makes the consideration of a measuring reference at $r=0$ meaningless.}. Hence, the concept of radar distance alone does not lead to consistent quantifications of $r$.

Students of gravitational physics may reasonably wonder how $m$ and $r$ could be found without using Newtonian or Euclidean approximations. The question stems from the idea that general relativity is an autonomous theory leading to self-contained applications; and should not require Newtonian gravity or Euclidean geometry to describe phenomena taking place in extended spacetime regions.

A comment found in page 140 of Foster and Nightingale's book \cite{fn}, regarding the indirect measurement of radial coordinate based on the knowledge of orbital period of a particle in circular motion, triggered our interest in this matter. According to the authors, the procedure requires the value of $M$, "which must be known independently, that is, not found by methods involving orbital periods." We believe that our approaches to Schwarzschild spacetime prove otherwise, and that $m$ and $r$ can be simultaneously determined from idealised observations of geodesic motion.

For a start, we note that the Schwarzschild metric implies simple relations between $m$, $r$ and various measurements of proper time. As a result, quantifications of mass parameter and orbital radius from experimental data should involve solutions to systems of algebraic equations.

The paper is organised as follows. In Section 2 we review circular orbits in Schwarzschild geometry and their stability properties, and consider measurements of orbital period by static observers. In Section 3 we propose thought experiments leading to indirect quantifications of mass parameter, orbital radius and radius of the central body, as well as more general determinations of radial marker through radar sounding. The pedagogical implications of our results are discussed in Section 4. An alternative, idealised procedure for measurement of mass parameter and orbital radius, motivated by the study of nearly circular orbits, is put forward in the Appendix. The material is primarily intended for advanced undergraduate or first year graduate students in physics. Our treatment of circular orbits draws mainly from \cite{fn}.

\section{Circular orbits}
The exact, static and spherically symmetric solution to the vacuum field equations reads (in $c=1$ units)
\begin{equation}\label{schm}
  ds^2=\left(1-\frac{2m}{r}\right)dt^2-\left(1-\frac{2m}{r}\right)^{-1}dr^2-r^{2}d\theta^2-r^2\sin^{2}\theta\,d\phi^2,
\end{equation}
where $t,r,\theta,\phi$ are the standard Schwarzschild coordinates \cite{fn}. The mass parameter $m$ is an integration constant with the dimension of length which takes positive values to conform to attractive gravitational effects\footnote{Adapting the geodesic equation to the condition of initial rest in Schwarzschild coordinates, and using the unit norm of the four-velocity for time-like worldlines, we find that the subsequent motion of a massive test particle is oriented towards the central body if and only if $m>0$. See also the footnote following (\ref{eqmr}).}. The assumption $r>2m$ guarantees the time-like character of $t$ and the space-like character of $r$. In the case of time-like paths, characterised by $ds^2>0$, the proper time $\tau$ associated with each test particle is invariantly defined by $ds^2=d\tau^2$.

The dynamics of time-like geodesics can be derived from the lagrangian
\begin{equation}\label{lagr}
  2\mathcal{L}=\left(1-\frac{2m}{r}\right)\dot{t}^2-\left(1-\frac{2m}{r}\right)^{-1}\dot{r}^2-r^{2}\dot{\theta}^2-r^2\sin^{2}\theta\,\dot{\phi}^2,
\end{equation}
where dots denote derivatives with respect to $\tau$. The equations of motion have the general form
\begin{equation}\label{gfe}
  \frac{d}{d\tau}\left(\frac{\partial \mathcal{L}}{\partial \dot{x}^{\mu}}\right)-\frac{\partial \mathcal{L}}{\partial x^{\mu}}=0,
\end{equation}
where the Greek index runs from $0$ to $3$, and $x^0 \equiv t$, $x^1 \equiv r$, $x^2 \equiv \theta$, $x^3 \equiv \phi$. The condition
\begin{equation}\label{tlc}
  \left(1-\frac{2m}{r}\right)\dot{t}^2-\left(1-\frac{2m}{r}\right)^{-1}\dot{r}^2-r^{2}\dot{\theta}^2-r^2\sin^{2}\theta\,\dot{\phi}^2=1
\end{equation}
establishes the time-like character of geodesic paths.

Combining (\ref{lagr}) and (\ref{gfe}) we get two first integrals and two second order equations, namely
\begin{align}
\label{leq0}
&\left(1-\frac{2m}{r}\right)\dot{t}=e, \\
\label{leq1}
&\left(1-\frac{2m}{r}\right)^{-1}\ddot{r}+\frac{m}{r^2}\,\dot{t}^2-\left(1-\frac{2m}{r}\right)^{-2}\frac{m}{r^2}\,\dot{r}^2-r\left(\dot{\theta}^2+\sin^{2}\theta \,\dot{\phi}^2\right)=0, \\
\label{leq2}
&\frac{d}{d\tau}\left(r^2\dot{\theta}\right)-r^2\sin\theta\cos\theta\,\dot{\phi}^2=0, \\
\label{leq3}
&r^2\sin^{2}\theta\,\dot{\phi}=l,
\end{align}
where $e$ and $l$ are integration constants.

Now we multiply (\ref{leq2}) by $r^2\dot{\theta}$, use (\ref{leq3}), integrate over $\tau$, and get the conservation law
\begin{equation} \label{nor1}
\left(r^2\dot{\theta}\right)^2+\frac{l^2}{\sin^{2}\theta}=L^2.
\end{equation}
where $L^2$ is a constant. We are free to choose spherical coordinates so that the initial values associated with the polar angle take the form $\theta(0)=\pi/2$, $\dot{\theta}(0)=0$. It follows that $L^2=l^2$, and the above equation reduces to
\begin{equation} \label{nor1}
\left(r^2\dot{\theta}\right)^2+l^2 \cot^{2}\theta=0.
\end{equation}
Clearly, each term in this equation is zero. Therefore, $\theta(\tau)=constant=\pi/2$ is the solution to the initial value problem \cite{nor}. In consequence, without loss of generality, (\ref{leq1}) and (\ref{leq3}) can be rewritten as
\begin{align}
\label{leq1n}
&\left(1-\frac{2m}{r}\right)^{-1}\ddot{r}+\frac{m}{r^2}\,\dot{t}^2-\left(1-\frac{2m}{r}\right)^{-2}\frac{m}{r^2}\,\dot{r}^2-r\dot{\phi}^2=0, \\
\label{leq3n}
&r^2\dot{\phi}=l.
\end{align}

Inserting the constant value of $\theta(\tau)$ into (\ref{tlc}), using the arising expression to eliminate $\dot{t}^2$ in (\ref{leq1n}), and taking (\ref{leq3n}) into account we get
\begin{equation}\label{eqmr}
\ddot{r}+\frac{m}{r^2}+\frac{l^2\left(3m-r\right)}{r^4}=0,
\end{equation}
which is the decoupled differential equation for the radial marker\footnote{A massive test particle initially at rest in Schwarzschild coordinates is characterised by $l=0$. The ensuing motion is radial and governed by $\ddot{r}=-m/r^2$, which admits the first integral $\dot{r}^2-2m/r=constant$ \cite{fn}. Clearly, $m>0$ is the necessary and sufficient condition for vertical fall (a feature of attractive gravity).}.

The assumption $r=a=constant$ combined with the above result leads to
\begin{equation}\label{l2f}
l^2=\frac{ma^2}{a-3m}\equiv l^2_0.
\end{equation}
Therefore, $l^2_0$ is positive only if $a>3m$, which is a necessary condition for the existence of time-like circular orbits.

Let us perturb a circular orbit without changing the hyperplane of motion ($\theta=\pi/2$) or the value of $l$. Ideally, this is achieved by applying momentary thrust in the radial direction so that $r$ and $\dot{\phi}$ remain essentially constant during the perturbation\footnote{The analogous Newtonian situation corresponds to an impulsive force acting along the radial direction, which exerts no torque with respect to the center of attraction, and does not change the angular momentum vector. We also note that more general perturbations of Schwarzschild orbits involving small initial changes in $r$ and $\dot{\phi}$ can leave $l$ unchanged provided that $2\delta r/r+\delta{\dot{\phi}}/\dot{\phi}=0$, as a consequence of (\ref{leq3n}).}. We set $r=a+\rho$ and $l=l_0$ in (\ref{eqmr}), and assume $|\rho|<<a$. Using (\ref{l2f}) and the binomial approximation $(1+\xi)^n\approx 1+n\xi$ ($|\xi|<<1$) we derive the linearised radial equation
\begin{equation}\label{aem}
\ddot{\rho}+\omega^2\rho=0,
\end{equation}
where
\begin{equation}\label{omega2}
\omega^2=\frac{m\left(a-6m\right)}{a^3\left(a-3m\right)}.
\end{equation}
If $a>6m$, then $\omega^2>0$ and (\ref{aem}) guarantees the stability of circular motion under small perturbations. On the other hand, orbits with $3m<a<6m$ are unstable. Interestingly, if $a=6m$, then $\omega^2=0$ and we need a higher-order binomial approximation to elucidate the dynamical situation. Plugging $r=6m+\rho$ and $a=6m$ into (\ref{eqmr}) and (\ref{l2f}), respectively, we find
\begin{equation}\label{rho2a}
\ddot{\rho}+\frac{\rho^2}{1296 \, m^3}=0.
\end{equation}
Then an arbitrarily small fluctuation $\rho$ around $a=6m$ entails $\ddot{\rho}<0$ regardless of the sign of $\rho$. Consequently, the response to the radial perturbation is not oscillatory and the involved "boundary orbit" is unstable\footnote{See \cite{snkl} for a detailed discussion of this type of orbit instability.}.

Hereinafter, we shall assume the existence of a stable circular orbit. Equivalently, the inequality $a>6m$ will be taken for granted in the upcoming derivations.

The choice $\theta=\pi/2$, $r=a=constant$ fixes the value of $\dot{\phi}$ as a consequence of (\ref{leq3n}) and (\ref{l2f}). We obtain
\begin{equation} \label{dppt2}
\dot{\phi}^2=\frac{m}{a^2(a-3m)}.
\end{equation}
The combination of this result with (\ref{leq1n}) leads to
\begin{equation} \label{dtpt2}
\dot{t}^2=\frac{a}{a-3m}.
\end{equation}
This expression and (\ref{leq0}) imply
\begin{equation}\label{efor}
e=\frac{a-2m}{\sqrt{a(a-3m)}},
\end{equation}
which allows the calculation of the additional conserved quantity in the case of a circular geodesic.

Using the property $d\phi/dt=\dot{\phi}/\dot{t}$ in conjunction with (\ref{dppt2}) and (\ref{dtpt2}) we get
\begin{equation} \label{cav}
\left(\frac{d\phi}{dt}\right)^2=\frac{m}{a^3}.
\end{equation}
This result provides a formula for the variation of $t$ in one cycle of the motion, given by
\begin{equation} \label{cav}
\Delta t=2\pi\,\sqrt{\frac{a^3}{m}}.
\end{equation}

Now we consider events $E$ and $R$ representing the emission and reception of a light signal, respectively. An important feature of the Schwarzschild geometry is that the coordinate time difference $t_R-t_E$ depends only on the spatial path associated with the null curve joining the events \cite{fn}. As a result, if the emitter and receiver are spatially fixed (in terms of Schwarzschild coordinates) and two successive signals are taken into account, we obtain
\begin{equation} \label{ipr}
\Delta t_R=t^{(2)}_R-t^{(1)}_R=t^{(2)}_E-t^{(1)}_E=\Delta t_E,
\end{equation}
which means that the coordinate time differences between consecutive emissions and consecutive receptions are the same. Let us assume that the first and second emissions correspond to the passage of the particle in circular motion through a certain spatial point, so that $\Delta t_E$ equals the period $\Delta t$. Hence, the proper time interval between receptions by a static observer, $\Delta\tau_R$, is given by
\begin{equation} \label{taut}
\Delta \tau_R=\sqrt{1-\frac{2m}{r_R}}\,\Delta t_R=\sqrt{1-\frac{2m}{r_R}}\,\Delta t=2\pi\,\sqrt{\left(1-\frac{2m}{r_R}\right)\frac{a^3}{m}},
\end{equation}
where $r_R>2m$ is the radial coordinate of the reception events, and use has been made of (\ref{cav}). This expression gives $\Delta \tau_R$ as a function of the unknown parameters $m$, $a$ and $r_R$.

\section{Thought experiments}
At this point we model a spaceship carrying two astronaut-observers as a test particle in circular motion around the central body. The first experiment begins when one of the astronauts leaves the craft, employs thrust to fix his/her position relative to the Schwarzschild coordinates, and remains as close to the circular orbit as possible. Then the static, hovering observer receives two successive light signals from his/her orbiting companion (emitted at the same fixed spatial point), measures the corresponding proper time interval $\Delta \tau_{hov}$, and safely returns to the spaceship. Inserting $r_R=a$ in (\ref{taut}) we get the formula
\begin{equation} \label{tauh}
\Delta \tau_{hov}=2\pi\,\sqrt{\left(1-\frac{2m}{a}\right)\frac{a^3}{m}}.
\end{equation}

In the second experiment, the astronauts employ an onboard clock to quantify the proper time elapsed during one cycle of the circular orbit, $\Delta \tau_{orb}$, which is also related to $m$ and $a$. From (\ref{dppt2}) we find
\begin{equation} \label{rtrp2}
\Delta \tau_{orb}=2\pi \sqrt{\left(1-\frac{3m}{a}\right)\frac{a^3}{m}}.
\end{equation}

The above results entail
\begin{equation} \label{tho}
\frac{\Delta \tau_{hov}}{\Delta \tau_{orb}}=\sqrt{\frac{a-2m}{a-3m}},
\end{equation}
and we notice that the double inequality $1 < \Delta \tau_{hov} / \Delta \tau_{orb} < 2/ \sqrt{3}$ holds.

Solving (\ref{tho}) for $\chi=a/2m>3/2$ we get
\begin{equation} \label{chi}
\chi=\frac{\frac{3}{2}\left(\Delta \tau_{hov}\right)^2-\left(\Delta \tau_{orb}\right)^2}{\left(\Delta \tau_{hov}\right)^2-\left(\Delta \tau_{orb}\right)^2}.
\end{equation}
In this manner, $\chi$ is determined from measurable proper time intervals.

Expressions (\ref{tauh}) or (\ref{rtrp2}) can be used to write $a$ as a function of $\chi$ and one of the proper time intervals. We arbitrarily choose the latter option and find
\begin{align}
\label{eqm1}
&m=\frac{\Delta \tau_{orb}}{4\sqrt{2}\,\pi\,\chi\sqrt{\chi-3/2}}, \\
\label{eqr1}
&a=\frac{\Delta \tau_{orb}}{2\sqrt{2}\,\pi\sqrt{\chi-3/2}}.
\end{align}
These results provide the values of mass parameter and orbit radius corresponding to $\Delta \tau_{hov}$ and $\Delta \tau_{orb}$.

In the third experiment various astronauts (carried by other spacecrafts) are positioned as static observers and receive the periodic signals emitted by the first vehicle. An astronaut at $r_R>2m$ quantifies the proper time interval $\Delta \tau_R$. Solving (\ref{taut}) for $r_R$ we get
\begin{equation} \label{rmx}
r_R=\frac{8\pi^2ma^3}{4\pi^2a^3-m\left(\Delta\tau_R\right)^2}.
\end{equation}
This formula yields the radial coordinate of the static observer as a function of the proper time elapsed between the arrivals of successive signals, and the indirectly measured values of $m$ and $a$. Particularly, if the receiver measuring $\Delta\tau_R$ is set on the surface of the central body, then (\ref{rmx}) gives its coordinate radius.

Radar sounding involves the round trip of an electromagnetic signal propagating in vacuum between a static observer and a reflection event. To specify the Schwarzschild line element for radial null paths, we set $ds^2=0$ as well as constant $\theta$ and $\phi$ in (\ref{schm}). This entails a relationship between the infinitesimal variations of $t$ and $r$ for events connected by a null ray, which can be integrated over $r$ to get the variation of coordinate time $\Delta t$ corresponding to the round trip. If the emitter-receiver is placed on the surface of the central body of radius $r_R>2m$, and $\Delta t$ is converted to proper time measured by the static observer we find, after integration by parts, the expression \cite{fn}
\begin{equation} \label{Dtau}
\Delta\tau=2\sqrt{1-\frac{2m}{r_R}}\,\left[r_e-r_R+2m\,\ln \left(\frac{r_e-2m}{r_R-2m}\right)\right],
\end{equation}
where $r_e>r_R$ is the radial marker of the reflection event. $\Delta\tau$ is twice the radar distance in $c=1$ units. We thus obtain
a transcendental equation that can be numerically solved for $r_e$ when $\Delta\tau$, $m$ and $r_R$ are known.

If $m$ and $a$ are quantified, then (\ref{rmx}) can be used to find radial marker values $r_1$ and $r_2$ for every pair of static astronaut-observers capable of measuring the proper time interval between the arrivals of two successive signals emitted by the orbiting spacecraft. As a result, the observers could send electromagnetic signals to each other and use the well-known formula
\begin{equation}\label{fse}
\frac{\nu_2}{\nu_1}=\sqrt{\frac{1-2m/r_1}{1-2m/r_2}}
\end{equation}
to explain gravitational spectral shifts \cite{fn} without resorting to the Newtonian limit or using Euclidean geometry to approximate the values of $r$.

\section{Concluding remarks}
General relativity and Newtonian gravity put forward very different conceptualisations of spacetime and gravitational interactions. These two frameworks are traditionally related through the Newtonian limit, so that the latter is seen as an approximation to the former. Actually, Newtonian gravity is more than that: in the case of the Schwarzschild solution, it allows the evaluation of $m$ in terms of the Newtonian gravitational mass. It appears, then, that general relativity's explanatory power would be hampered if the limit was ignored. Such dependence of Einstein's theory on Newton's gravity is likely to puzzle students, who may end up seeing general relativity as an incomplete physical theory. Our analysis of circular orbits aims to strengthen the autonomy of the Schwarzschild geometry, avoiding the difficult asymptotic identification of curvature coordinates with Newtonian coordinates. This purpose is meaningful even if we restrict our attention to thought experiments.

We have shown that formulas derived for the Schwarzschild spacetime can be used in a self-contained manner, i.e. without the support of Newtonian gravity in asymptotic regions, or approximations involving Euclidean geometry. Beyond that, applications of these ideas to static spacetimes lacking asymptotic flatness are also possible. In those cases the implementation of the Newtonian limit can be more difficult, or even impossible.

Students interested in the conceptual structure of general relativity may go further and use the above discussions to reconsider the meaning of $M$, which is a measurable quantity in Newton's gravitational theory. Imported into Einstein's theory through the implementation of the Newtonian limit, it appears in the Schwarzschild metric. Its numerical value (obtained by means of Newtonian and Euclidean procedures) affects general relativistic calculations. The question arises as to how $M$ could be defined in the language of Einstein's theory, where gravitational forces do not exist. In the absence of a consistent definition we couldn't help but see it as a foreign parameter without a clear content. The query is left open.

Motivated students may also examine coordinate choices in general relativity. According to conventional wisdom, derived from the principle of general covariance, "coordinates are simply labels of spacetime events that can be assigned completely arbitrarily (...). The only quantities that have physical meaning \hspace{0.1cm} -the measurables- \hspace{0.1cm} are those that are invariant under coordinate transformations. One such invariant is the number of ticks on an atomic clock giving the proper time between two events."\cite{fpw}. This observation is highly relevant to the present work. In particular, our analysis suggests that the Schwarzschild coordinate $r$ can be ultimately expressed as a function of various measurements of proper time -each one recognised as an invariant quantity with physical meaning. The inevitable questions follow: Should $r$ itself be considered as a physical quantity? What about "the fact that the Schwarzschild geometry can be described in infinitely many coordinate systems"?\cite{fpw}. Should each of the alternative, infinitely many radial coordinates be regarded as an observable? And what can be said about the Schwarzschild coordinate $t$? Would it end up being another physical quantity, as suggested by (\ref{cav}), given that $m$ and $a$ emerge as functions of proper time intervals? How to consider such possibilities in light of the principle of general covariance? These queries are left open as well.

\renewcommand{\theequation}{A\arabic{equation}}
\setcounter{equation}{0}

\section*{Appendix}
A different type of thought experiment combines the determination of $m$ and $r$ with the study of nearly circular orbits. Let us imagine that, after measuring $\Delta\tau_{orb}$, the spacecraft's crew briefly employs thrust to slightly perturb their stable circular orbit of radius $a$ while leaving $l$ constant. To discuss this situation we solve (\ref{aem}) and obtain
\begin{equation}\label{rtau}
r(\tau)=a+A\cos(\omega\tau)+B\sin(\omega\tau),
\end{equation}
where $A$ and $B$ are integration constants. Therefore, the radial marker oscillates about its mean value $a$ with angular frequency $\omega$. Using (\ref{omega2}) we find the corresponding period
\begin{equation} \label{pp}
T=2\pi\,\sqrt{\frac{a^3(a-3m)}{m(a-6m)}}.
\end{equation}

We assume that an onboard gravity gradiometer detects variations of tidal force caused by very small changes in spatial separation between the spacecraft and the central body\footnote{It is worth pointing out that a very sensitive, fast electrostatic gravity gradiometer went into orbit in 2009 as part of the European Space Agency's GOCE mission. See \cite{goce,or} for more details.}. This would enable the astronaut-observers to quantify the proper time $T$ elapsed between two successive maximum (or minimum) separation events. From (\ref{rtrp2}) and (\ref{pp}) we get
\begin{equation}\label{msap}
\frac{m}{a}=\frac{1}{6}\left[1-\left(\frac{\Delta\tau_{orb}}{T}\right)^2\right],
\end{equation}
which implies $T>\Delta\tau_{orb}$. Also,
\begin{align}
\label{mper}
&m=\frac{\Delta\tau_{orb}}{12\sqrt{3}\,\pi}\,\frac{\left[1-\left(\Delta\tau_{orb}/T\right)^2\right]^{3/2}}{\left[1+\left(\Delta\tau_{orb}/T\right)^2\right]^{1/2}},\\
\label{aper}
&a=\frac{\Delta\tau_{orb}}{2\sqrt{3}\,\pi}\,\left[\frac{1-\left(\Delta\tau_{orb}/T\right)^2}{1+\left(\Delta\tau_{orb}/T\right)^2}\right]^{1/2}.
\end{align}

To complete our description of the perturbed motion, we impose the initial condition $\dot{r}(0)=0$ and reduce (\ref{rtau}) to
\begin{equation}\label{rtaur}
r(\tau)=a+A\cos(\omega\tau).
\end{equation}
Plugging this into (\ref{leq3n}), using $|\rho|<<a$, and invoking the binomial approximation we are led to
\begin{equation}\label{phip}
\dot{\phi}=\frac{l_0}{a^2}-\frac{2 l_0 A}{a^3}\, \cos(\omega\tau),
\end{equation}
so the total variation of $\phi$ in one cycle of the radial oscillation is given by
\begin{equation}\label{fphi}
\Delta\phi_{T}=\frac{l_0 T}{a^2}=\frac{2\pi}{\sqrt{1-6m/a}},
\end{equation}
where (\ref{l2f}) and (\ref{pp}) have been used. This gives the apsidal precession angle
\begin{equation}\label{epsq}
\epsilon=\Delta\phi_{T}-2\pi=2\pi\left[\frac{1}{\sqrt{1-6m/a}}-1\right].
\end{equation}
The weak gravity assumption $a>>6m$ leads to the simpler formula
\begin{equation}\label{epsa}
\epsilon=\frac{6\pi m}{a}.
\end{equation}
We note that, after minimal changes in notation, (\ref{fphi}) and (\ref{epsa}) reproduce (25.48) and (11.37) in \cite{mtw} and \cite{bs}, respectively.

The above results show that apsidal precession formulas for nearly circular orbits can be simply obtained from the linearized equations of motion without considering the explicit dependence $r=r(\phi)$.  Also, (\ref{epsq}) and (\ref{epsa}) constitute predictions of apsidal precession made independently of the Newtonian limit of general relativity.

We have assumed that the value of $l$ is unaffected by a particular perturbation of circular motion. To describe the behaviour of the other conserved quantity, $e$, we eliminate $\dot{t}$ from (\ref{tlc}) and (\ref{leq0}), use (\ref{leq3n}), and find
\begin{equation}\label{seq}
e^2=\dot{r}^2+\left(1-\frac{2m}{r}\right)\left(1+\frac{l_{0}^2}{r^2}\right),
\end{equation}
where $l$ has been substituted by $l_0$. Plugging $r=a+\rho$ into the above expression, expanding the arising function of $\rho$ up to and including the term in $\left(\rho/a\right)^2$, and using (\ref{l2f}), (\ref{omega2}) and (\ref{efor}) we get
\begin{equation}\label{erho}
e^2=e_{0}^2+\dot{\rho}^2+\omega^2\rho^2+O\left((\rho/a)^3\right).
\end{equation}
Here $e_0$ is the value of the conserved quantity before perturbation. From (\ref{rtaur}) we see that $\rho=A\cos(\omega \tau)$, so (\ref{erho}) takes the form
\begin{equation}\label{econs}
e^2=e_{0}^2+\omega^2 A^2+O\left((\rho/a)^3\right),
\end{equation}
where the sum of the first two terms on the right hand side gives the constant value of $e^2$ in the linear approximation of the dynamics\footnote{We highlight that the second term on the right hand side of (\ref{seq}) is the effective potential for time-like orbits (massive particles) of fixed $l=l_0$ \cite{bs}. In graphical form, constant quantities $e^{2}_0$ and $e^2$ are represented by horizontal lines superimposed upon the effective potential curve. The three relevant intersections correspond to the radius of the initial circular orbit, $a$, and the minimum and maximum values of $r$ for the precessing orbit. Perturbations that leave $l$ constant involve transitions between energies $e_0$ and $e$ that do not change the shape of the effective potential.}.

Provided that $e_0$ and $\omega$ are functions of the indirectly measurable quantities $m$ and $a$, we just need to quantify $A$ to find $e$. To this end, we assume that the maximum and minimum separations between the satellite and the surface of the central body can be determined through radar sounding, and use (\ref{Dtau}) to relate $A$ to the largest variation of the observable quantity $\Delta \tau$.

Substituting $r_e=a+A$ and $r_e=a-A$ into (\ref{Dtau}) we respectively get expressions for the maximum and minimum values of $\Delta\tau$, given by
\begin{align}
\label{Dtx}
&\Delta\tau_{max}=2\sqrt{1-\frac{2m}{r_R}}\,\left[a+A-r_R+2m\,\ln \left(\frac{a+A-2m}{r_R-2m}\right)\right],\\
\label{Dtn}
&\Delta\tau_{min}=2\sqrt{1-\frac{2m}{r_R}}\,\left[a-A-r_R+2m\,\ln \left(\frac{a-A-2m}{r_R-2m}\right)\right].
\end{align}
Now we treat the difference $\Delta\tau_{max}-\Delta\tau_{min}$ as a function of $A$ and obtain its Maclaurin series up to and including the term in $A$, namely
\begin{equation}\label{dDxm}
\Delta\tau_{max}-\Delta\tau_{min}=4\sqrt{1-\frac{2m}{r_R}}\,\left[\frac{A}{1-2m/a}+O(A^3)\right].
\end{equation}
The first order term in this expansion entails a formula for $A$ as a function of $m$, $a$, $r_R$, $\Delta\tau_{max}$ and $\Delta\tau_{min}$, which combines with (\ref{econs}) and ultimately gives the approximate value of $e$.

We point out that (\ref{epsq}) or (\ref{epsa}) can be solved with (\ref{rtrp2}) to get $m$ and $a$ in terms of $\Delta \tau_{orb}$ and apsidal precession angle $\epsilon$. Similarly, the geodesic effect \cite{fn} provides an expression for the precession angle of an orbiting spinning gyroscope that can be combined with (\ref{rtrp2}) to find $m$ and $a$ as functions of $\Delta \tau_{orb}$ and the foregoing angle. We leave the analysis of thought experiments involving angular measurements for future work\footnote{It should be noted that the actual satellite-based measurement of the geodesic effect in the Earth's gravitational field was accomplished as part of the Gravity Probe B experiment \cite{stnfd,evrt,cmw}.}.

\end{document}